\documentclass[aps,prb,twocolumn,superscriptaddress]{revtex4-1}

\usepackage{graphicx}
\usepackage{amsmath}
\usepackage{amstext}
\usepackage{graphicx}
\usepackage{amsmath}
\usepackage{amstext}
\usepackage{amssymb}
\usepackage{amsfonts}
\usepackage{amsbsy}
\usepackage{dcolumn}
\usepackage{bm}
\usepackage{esint}
\usepackage{multirow}
\usepackage{hyperref}
\hypersetup{
    colorlinks=true,
    linkcolor=blue,
    filecolor=magenta,
    urlcolor=cyan,
}
\usepackage{cleveref}

\usepackage{mathrsfs}
\usepackage{amsfonts}
\usepackage{amsbsy}
\usepackage{dcolumn}
\usepackage{bm}
\usepackage{multirow}
\usepackage{color}
\def\ep{{\epsilon^{\mu\nu\lambda\rho} }}
  \usepackage{extarrows}

\usepackage{datetime}

\begin{document}
 \title{{\fontfamily{ptm}\fontseries{b}\selectfont Topological quantum field theory of three-dimensional bosonic Abelian-symmetry-protected topological phases}}

  \author{Peng Ye}
      \affiliation{Department of Physics and Institute for Condensed Matter Theory,
University of Illinois at Urbana-Champaign, IL 61801, USA}
    \affiliation{Perimeter Institute for Theoretical Physics, Waterloo, Ontario, Canada N2L 2Y5}

    \author{Zheng-Cheng Gu}
   \affiliation{ Department of Physics, The Chinese University of Hong Kong, Shatin, New Territories, Hong Kong}
  \affiliation{Perimeter Institute for Theoretical Physics, Waterloo, Ontario, Canada N2L 2Y5}

\begin{abstract}
Symmetry-protected topological phases (SPT)   are short-range entangled gapped states protected by global symmetry.  Nontrivial SPT phases cannot be adiabatically connected to the trivial disordered state(or atomic insulator) as long as certain global symmetry $G$ is unbroken.  At low energies,  most of two-dimensional SPTs with Abelian symmetry can be described by topological quantum field theory (TQFT) of multi-component Chern-Simons type. However, in contrast  to the fractional quantum Hall effect where TQFT can give rise to interesting bulk anyons, TQFT for SPTs only supports trivial bulk excitations.  The essential question in TQFT descriptions for SPTs is to understand how the global symmetry is implemented in the partition function. In this paper, we systematically study TQFT of  three-dimensional SPTs with unitary Abelian symmetry (e.g., $\mathbb{Z}_{N_1}\times\mathbb{Z}_{N_2}\times\cdots$). In addition to the usual multi-component   $BF$ topological term at level-$1$, we find that there are new topological terms with quantized coefficients (e.g., $a^1\wedge a^2\wedge d a^2$ and $a^1\wedge a^2\wedge a^3\wedge a^4$) in TQFT actions, where $a^{1},a^2,\cdots$ are 1-form U(1) gauge fields.  These additional topological terms cannot be adiabatically turned off as long as $G$ is unbroken. By investigating symmetry transformations for the TQFT partition function, we end up with the classification of SPTs that is consistent with the well-known group cohomology approach.  We also discuss how to gauge the global symmetry and possible TQFT descriptions of Dijkgraaf-Witten gauge theory.  
 \end{abstract}
  \maketitle
 \section{Introduction}

 \subsection{Background and motivation}

At low energies, universal properties of quantum many-body systems may be governed by continuum quantum field theory. For symmetry-breaking phases, Ginzburg-Landau field theory with symmetry-breaking potentials is utilized, where the concept of local order parameters is introduced. 
For topological phases of quantum matter, such as fractional quantum Hall effects (FQHE), local order parameters vanish and thereby fail to  characterize the ground state. However, such phases can be described by   \emph{topological} quantum field theory (TQFT) \cite{chetan08,Wenbook}.  As a renormalization fixed point theory, TQFT partition functions are usually formulated in the Euclidean path-integral formalism where the classical action is
 purely imaginary   and all correlation functions of local operators are invariant under diffeomorphism of the spacetime manifold. In this sense, TQFTs are expected to  efficiently capture the global   phenomena that are insensitive to local energetic details.  For example, the (2+1)-dimensional  Chern-Simons field theory was introduced  to describe FQHE  \cite{chetan08,Wenbook,km1,km2,km3,km4,km5,km6,km7,composite1,jain,boson}.  It reveals the key mechanism of flux-charge attachment for FQHE and  successfully predicts the anyon statistics and quantized Hall conductance \cite{Wenbook}.
 Very recently, there is intensive ongoing research  focusing on the interplay between symmetry and topology in strongly correlated systems. For example, the so-called ``symmetry-protected topological phases'' (SPT) in interacting bosonic / spin systems \cite{1DSPT,Chenlong,Chen10,Wenscience,Wencoho,2d11,2d8,2d1,2d_pollmann,2dd,2d3,2d2,2d4,2d5,2d6,2d61,2d7,2d9,2d10,2d12,2d13,2d14,2d16,2d17,2d18,2d19,2d20, wangwen_twist,ran,fieldbi, 3looplong1,3looplong,wan3d,hung,corbodism1,corbodism2,corbodism3,wang_levin1,3d1,3d2,3d3,3d4,3d5,3d6,Geraedts,alicea,gaioto,yegu2014,WGW,GWW,abeffect}  have been drawing much attention.  In contrast to FQHE  where the bulk is fractionalized and supports topologically anyonic quasiparticles, the bulk of bosonic SPT  phase is non-fractionalized and only supports trivial  bosonic quasiparticles. Nevertheless, distinct SPT phases can still be identified as long as  a symmetry group, say, $G$ is unbroken. In other words, two distinct SPT phases with $G$  cannot be adiabatically connected to each other unless $G$ symmetry is broken\cite{Chenlong}.

Since TQFT approach has been successfully applied in topological phases of quantum matter (e.g., FQHE), a natural question would be:  Can we also use TQFT to describe universal properties of SPT phases that, by definition, do not support fractionalized bulk excitations? If so, we expect that such TQFTs must  have  two  key properties:
\begin{enumerate}
\item  Bulk excitations are non-fractionalized. In other words, they have a unique ground state on any closed manifold and there do not exist nontrivial topological sectors.  
\item TQFTs of distinct SPT phases are topologically distinguishable from each other  if and only if certain unbroken symmetry $G$ is considered.  It indicates that   nontrivial properties of TQFT are topologically robust only when $G$ symmetry is unbroken.
\end{enumerate}

In  Ref.~\onlinecite{2d8}, it was proposed to use multicomponent Abelian Chern-Simons theory  to describe 2D SPT  phases with Abelian symmetry. The   Chern-Simons theory applied in SPT phases must have a unique ground state on a torus. SPT phases are classified by distinct ways of anomalous symmetry transformations implemented on the 1D boundary.  In Ref.~\onlinecite{GWW}, a complete TQFT description for 2D SPT  phases with Abelian symmetry has been achieved.  
  {Nevertheless,   a complete TQFT description  of 3D  SPT  phases is more challenging and still not conclusive even for Abelian symmetry groups.}  In   Ref.~\onlinecite{yegu2014}, bosonic topological insulators (BTI) as specific cases of 3D SPT phase are studied and the corresponding TQFTs are derived via the vortex-line condensation scenario, where the cosmological constant-type term ``$b\wedge b$'' is used. 
Ref.~\onlinecite{3d3} proposes to use topological ${BF}$ field theory to describe some SPT phases in the presence of $\mathrm{U(1)}$ charge conservation symmetry.
But except these special cases,  we actually know very little about the TQFT description of  3D SPT, especially with unitary discrete symmetry.   Indeed,   there have already been some interesting proposals of 3D SPT materials, such as  ``topological paramagnetism'' in 3D  integer-spin Mott insulators \cite{WSenthil_topopara}.     We hope that a full establishment of TQFT in every spatial dimension will   help us to systematically identify   low-energy properties of bosonic SPT phases and realize them in a lab in future.

 \begin{figure}[t]
\centering
\includegraphics[width=8cm]{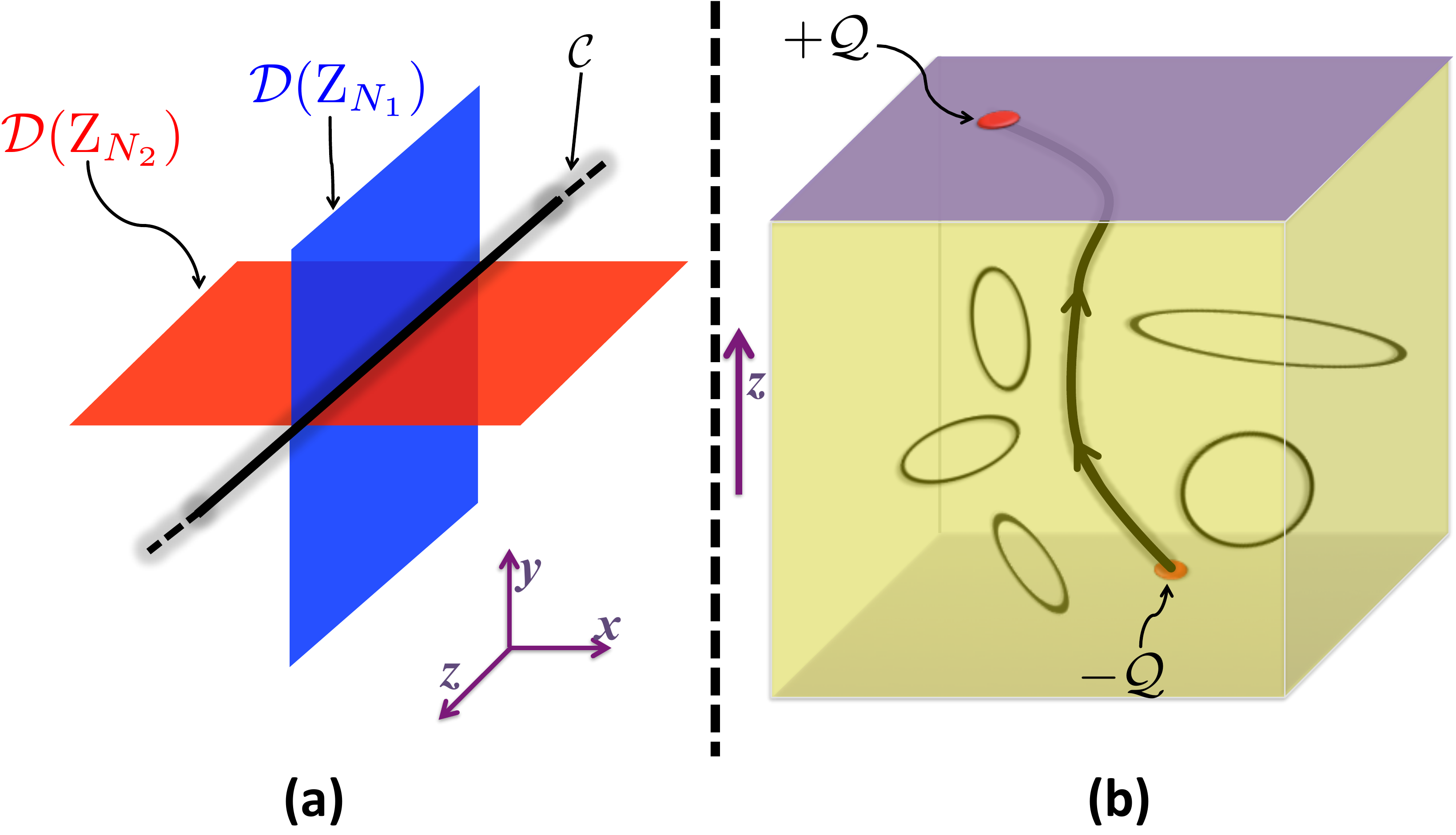}
\caption{(Color online)   {Bosonic wormhole effect  as a physical mechanism of  SPT  phases with $\prod^2_I \mathbb{Z}_{N_I}$ symmetry}. \textbf{(a)} A $\mathbb{Z}_{N_1}$ symmetry domain wall ``$\mathcal{D}(\mathbb{Z}_{N_1})$'' and a $\mathbb{Z}_{N_2}$ symmetry domain wall ``$\mathcal{D}(\mathbb{Z}_{N_2})$'' intersect along a closed loop ``$\mathcal{C}$'' (only a segment is shown due to the limited space).
\textbf{(b)} shows the bosonic wormhole effect induced by intersections of symmetry domain walls. Each intercepting line forms a closed loop in the bulk,  but some of them may form open strings and end at two 2D  boundaries between which the bulk is sandwiched.  Along such an open string $\mathcal{C}$ (the line with arrow),  $\mathbb{Z}_{N_2}$ symmetry charge $\mathcal{Q}$ is pumped from the endpoint on the lower boundary to the endpoint on the upper boundary.  See main text for more detailed explanation.}
\label{figure_aada}
\end{figure}

\subsection{Main results and outline}

In this paper, we   attempt to establish a TQFT description of   3D SPT phases with  unitary Abelian symmetry (see Table \ref{table1}).  Instead of  the abstract group cohomology approach \cite{Chenlong,Wencoho}, the  strategy we adopt in this paper is based on Ginzburg-Landau (GL) type actions with global symmetry (Sec.~\ref{sec:GLSPT}). Such an approach is more  physical and accessible.  The physical mechanism for realizing SPT phases  is pictorially illustrated in  Fig. \ref{figure_aada},  where the  ``{bosonic wormhole effect}'' is introduced. Then,  by dualizing  GL actions, we obtain   TQFT partition functions (Sec.~\ref{sec:TQFTSPT}).  We quantitatively study how symmetry transformations are realized in TQFTs, which gives rise  to the same SPT classification results that were previously obtained from group cohomology approach  \cite{Chenlong}. While  SPT phases with $\prod^2_I\mathbb{Z}_{N_I}$ are studied in details,  other SPT phases can be understood in a similar manner (Sec.~\ref{sec:general_symmetry}), which are summarized in Table \ref{table1}.  As such, TQFTs in this paper may be regarded as continuum field theory of group cohomological lattice models.  We  also  connect our TQFT results to the ``decorated domain wall picture''  \cite{2d19} as well as the interesting `3-loop statistics' phenomena \cite{wang_levin1} (Sec.~\ref{sec:general_symmetry}).  
A brief summary of the paper including some future directions is given in Sec.~\ref{sec:conclusions}. Several technical details are collected in Appendices.

\begin{table}
\caption{A brief summary of \textit{irreducible} 3D SPT phases with unitary Abelian symmetry. By ``irreducible'', we means that all subgroups of symmetry group play nontrivial roles in protecting the nontrivial SPT phases.   $a^I$  and $b^I$ are 1-form and 2-form $\mathrm{U(1)}$ gauge fields, respectively. 
``($\mathbb{Z}_{N_{12}}$)$\cdots$'' denote the corresponding classifications, where $N_{IJ\cdots}$ are greatest common divisors of $N_I, N_J,\cdots$.
SPT  phases with either $\mathbb{Z}_N$ or $\mathrm{U(1)}^k$ or $\mathbb{Z}_N\times\mathrm{U(1)}^k$ are trivial and not included below. All other  SPT's with unitary Abelian group symmetries can be obtained directly by using this table. For example, with arbitrary unitary finite Abelian group $G=\mathbb{Z}_{N_1}\times \mathbb{Z}_{N_2}\cdots$, it is sufficient to use this table to reproduce all SPT phases classified by the group cohomology, since
$\mathcal{H}^{4}(\mathbb{Z}_{N_1}\times \mathbb{Z}_{N_2}\cdots,\mathrm{U(1)})=\Pi_{I<J}(\mathbb{Z}_{N_{IJ}})^2\!\times\!\Pi_{I<J<K}(\mathbb{Z}_{N_{IJK}})^2\!\times\!\Pi_{I<J<K<L}\mathbb{Z}_{N_{IJKL}}$. }\label{table1}
 \begin{tabular}[t]{cc}
\hline

\hline
\hline
 \begin{minipage}[t]{0.6in}\textbf{Symmetry Group}\end{minipage} &\begin{minipage}[t]{2.7in} \textbf{Topological Quantum Field Theory and Classification}  \end{minipage}     \\\hline
   \begin{minipage}[t]{0.6in}$\mathbb{Z}_{N_1}\times\mathbb{Z}_{N_2}$ \end{minipage}&\begin{minipage}[t]{2.7in}
    $\frac{i}{2\pi}\int \sum^2_Ib^I\wedge \,da^I+i  p_1\int a^1\wedge a^2\wedge \,da^2$  ($\mathbb{Z}_{N_{12}}$);\\    $\frac{i}{2\pi}\int \sum^2_Ib^I\wedge \,da^I+ip_2\int  a^2\wedge a^1\wedge \,da^1$  $(\mathbb{Z}_{N_{12}}$)\end{minipage}\\\hline
   \begin{minipage}[t]{0.6in}$\mathbb{Z}_{N_1}\times\mathbb{Z}_{N_2}\times\mathbb{Z}_{N_3} $ \end{minipage}&\begin{minipage}[t]{2.7in}
    $\frac{i}{2\pi}\int \sum^3_Ib^I\wedge \,da^I+ip_1 \int a^1\wedge a^2\wedge \,da^3$ ($\mathbb{Z}_{N_{123}}$); \\$\frac{i}{2\pi}\int \sum^3_Ib^I\wedge \,da^I+ip_2 \int  a^2\wedge a^3\wedge \,da^1$ ($\mathbb{Z}_{N_{123}}$)\end{minipage}     \\\hline
   \begin{minipage}[t]{0.6in}$\prod^4_I\mathbb{Z}_{N_I} $  \end{minipage} &\begin{minipage}[t]{2.7in}  $\frac{i}{2\pi}\int \sum^4_Ib^I\wedge \,da^I+ip \int  a^1\wedge a^2\wedge a^3\wedge   a^4$ ($\mathbb{Z}_{N_{1234}}$) \end{minipage}   \\\hline
   \begin{minipage}[t]{0.6in}$\mathbb{Z}_{N_1}\times\mathbb{Z}_{N_2}\times \mathrm{U}(1)$  \end{minipage}&\begin{minipage}[t]{2.7in}  $\frac{i}{2\pi}\int \sum^3_Ib^I\wedge \,da^I+ip\int  a^1\wedge a^2\wedge \,da^3$ ($\mathbb{Z}_{N_{12}}$)\end{minipage}

     \\

        \hline

       \hline\hline

  \end{tabular}
  \end{table}

\section{SPT as a quantum disordered phase: Ginzburg-Landau action approach}\label{sec:GLSPT}

\subsection{A review of quantum disordered phase}
We start with a brief review of the quantum field theory description for quantum disordered phase. Let us consider the following partition function of four-dimensional classical XY model in Villain form:
\begin{align}
\mathcal{Z}=\int D\theta \sum_{\{N_{ij}\}}e^{-\frac{\chi}{2}\sum_{\langle ij\rangle}(\theta_i-\theta_j-2\pi N_{ij})^2} \,,\label{Villain}
\end{align}
where $N_{ij}\in\mathbb{Z}$ is an integer link variable defined on the NNN link $\langle ij\rangle$ of 4D cubic lattice.  $\chi$ is inverse temperature. $\theta_i\in (0,2\pi]$ is a $\mathrm{U(1)}$ phase angle defined on lattice site $i$. This classical model corresponds to a quantum XY model in $3+1$D spacetime.   There is a phase transition which separates two quantum phases at $\chi=\chi_c$. At small $\chi$ region near $\chi_c$ where the Villain approximation works, we have a ``quantum disordered'' phase where the ground state respects $\mathrm{U(1)}$ global symmetry and the lowest excitations are gapped. Such a quantum disordered phase serves as our starting point for understanding SPT phases.

To proceed further with field theory language, we rewrite the above lattice theory in terms of continuous spacetime variables with the following Lagrangian:
\begin{align}
\mathcal{L}_{0}=& \frac{\chi}{2}(\partial_\mu \theta)^2=\frac{\chi}{2}(\partial_\mu \theta_s\!+\!\partial_\mu \theta_v)^2
\!=\! \frac{\chi}{2}(\partial_\mu \theta_s-{a}_\mu)^2 ,\label{action0}
\end{align}
where we decompose $\theta$ as $\theta=\theta_s+\theta_v$. The $\theta_s$ variable in Eq.~(\ref{action0}) is a smooth function. The singular part(or vortex part) $\theta_v$ of original lattice model can be incorporated in the newly-introduced vector variable $a_\mu$.  It can be viewed as a $\mathrm{U(1)}$ gauge field since the action is manifestly invariant under $\theta\rightarrow \theta+ \chi, a_\mu\rightarrow a_\mu+\partial_\mu \chi$ where $\chi$ is a scalar gauge parameter. Apparently, such a gauge degree of freedom origins from the ambiguity(up to the scalar field $\chi$) of the decomposition $\theta=(\theta_s+\chi)+(\theta_v-\chi)$.

We note that the action Eq.~(\ref{action0}) actually only describes one of local minima of the Villain form Eq.~(\ref{Villain}). Since $\theta$ is by definition smooth, the large gauge transformation of $a$ with singular $\chi$ and $\int_{S^1}d\chi=2\pi n$ (with $n$ an arbitrary integer) will transform the action Eq.~(\ref{action0}) to another local minimum. Therefore, in order to maintain large gauge invariance, it is required to sum over all local minima, which essentially reproduce the Villain form Eq.~(\ref{Villain}). Thus, in the path integral formulation $\mathcal{Z}=\int D\theta  Da e^{-\int d^4 x \mathcal{L}_0}$, all the large gauge transformations of $a$ will be naturally included.

\subsection{Ginzburg-Landau action description for a $3+1$D bosonic SPT phase with $G=\mathbb{Z}_{N_1}\times \mathbb{Z}_{N_2}$ symmetry}
Since there are no interesting SPT phases with $ \mathbb{Z}_N$ or $\mathrm{U(1)}^k$ symmetry in $3+1$D, here we consider the simplest $3+1$D bosonic SPT phase with Abelian symmetry $\text{G}= \mathbb{Z}_{N_1}\times \mathbb{Z}_{N_2}$. 
We begin with the following Ginzburg-Landau action with $\mathrm{U(1)}\times \mathrm{U(1)}$ symmetry.
\begin{align}
S_{0}=&\sum^2_I\!\!\int \!\!d^4x\!\! \left[ \frac{\chi}{2}(\partial_\mu \theta^I_s-{a}_\mu^I)^2+\frac{1}{4g^2}(f^I_{\mu\nu})^2\right]+\cdots\,.\label{action01}
\end{align}
At the quadratic level of the low energy effective theory, we may also add a ``Maxwell'' term $\frac{1}{4g^2}(f^I_{\mu\nu})^2$ where $f_{\mu\nu}=\partial_\mu a_\nu-\partial_\nu a_\mu$ and $g$ is gauge coupling constant.

In addition to $S_0$, we may add the following topological term:
\begin{align}
S_{\rm Top} \!=\!-i p\int \!\! d^4x\ep(\partial_\mu\theta^1_s-a^1_\mu)(\partial_\nu\theta^2_s-a^2_\nu)\partial_\lambda  a^2_\rho.\label{topo_interaction123}
\end{align}
Mathematically, $S_{\rm Top}$ is a wedge product of differential forms that is invariant under diffeomorphism. $S_{\rm Top}$ is the only term in (3+1)D spacetime that is topological and gauge-invariant with two independent 1-form gauge fields.  Although $S_{\rm Top}$ formally has a $\mathrm{U(1)}\times \mathrm{U(1)}$ symmetry, below we will show that  such a symmetry is anomalous and can not be realized in a strictly 3D system. However, when $p$ takes certain quantized value, Eq.(\ref{topo_interaction123}) has an anomaly-free $\mathbb{Z}_{N_1}\times \mathbb{Z}_{N_2}$ symmetry. In Sec. \ref{sec:TQFTSPT}, we will rigorously show that   $p$ is not only quantized but also takes values in a compact space. Distinct SPTs may be labeled by distinct $p$'s, which  may unveil the classifications of SPT phases.  As a matter of fact, a usual dimension reduction scheme may already provide us a simple picture about the quantization condition on $p$. 

For this purpose, let us consider a special $4$D manifold $\mathcal{M}^4=T^2\times T^2$, where $\int_{T^2} d a^2=2\pi $ on one of the torus $T^2$. After compactifying over this $T^2$, we end up with a $1+1$D
SPT phases described by the following topological term:
\begin{align}
S_{\rm Top}^\prime \!=\!-2\pi i {p}\int \!\! d^2x\epsilon^{\mu\nu}(\partial_\mu\theta^1_s-a^1_\mu)(\partial_\nu\theta^2_s-a^2_\nu) \,.
\end{align}
In Ref.~\onlinecite{GWW}, it has been shown that if $4\pi^2 p=0$ mod $1$, the above topological term has an anomalous $\mathrm{U(1)}\times \mathrm{U(1)}$ symmetry and can not be realized as a pure $1+1$D system. However, if $4\pi^2 p=0$ mod $\frac{N_1N_2}{N_{12}}$ with  $N_{12}$ the greatest common divisor of  $N_1$ and $N_2$, the above topological term has an anomaly-free $ \mathbb{Z}_{N_1}\times  \mathbb{Z}_{N_2}$ symmetry. Therefore, if $p=\frac{N_1N_2 k}{4\pi^2 N_{12}}$ where $k=0,1,\cdots N_{12}$, the original topological term Eq.~(\ref{topo_interaction123}) will describe ${N_{12}}$ different $3+1$D SPT phases with $ \mathbb{Z}_{N_1}\times  \mathbb{Z}_{N_2}$ symmetry. In Sec.~\ref{sec:TQFTSPT}, the classification will be systematically derived from a TQFT which is a dual theory of the above GL action.

  The physical meaning of $S_{\rm Top}$ can be understood via {bosonic wormhole effect} induced by intersections of symmetry domain walls.  On one hand, the wormhole effect of  topological insulators was discussed in Ref.~\onlinecite{franz}, where the bulk is a fermionic system. On the other hand, by definition,
a $\mathbb{Z}_{N_I}$ symmetry domain wall appears as a  2D  closed manifold that separates two 3D regions with two distinct $\theta^I=\frac{2\pi}{N_I}k_I$ with $k_I\in\mathbb{Z}_{N_I}$.   We consider $a^1_\mu=a^2_\mu=0$ configurations such that $\mathbb{Z}_{N_I}$ symmetry domain walls ($I=1,2$) do not proliferate. As a result, there are many closed loops that are intersections between $\mathbb{Z}_{N_1}$ and $\mathbb{Z}_{N_2}$ symmetry domain walls as shown in Fig.\ref{figure_aada}-(a). Such closed loops do not contribute a $\mathbb{Z}_{N_2}$ symmetry charge $\mathcal{Q}$ from $S_{\rm Top}$ since $\mathcal{Q}= \partial_i (\partial_j\theta^1\partial_k\theta^2)\epsilon^{ijk}=0$.  However, if open boundaries are considered,  intersections may form open strings that end at two 2D boundaries as shown in Fig.\ref{figure_aada}-(b). At these endpoints,  nonzero $  \mathcal{Q}$ is realized as a pumped charge due to wormhole effect. Each symmetry domain wall is labeled by $\frac{2\pi K_I}{N_1}$, where $K_I\in\mathbb{Z}_{N_I}$ $(I=1,2)$. As such, $\mathcal{Q}=  p\frac{2\pi K_1}{N_1 }\frac{2\pi K_2}{N_2}=  \frac{k \,K_1K_2}{N_{12}}$ ($k\in\mathbb{Z}_{N_{12}}$) where the quantization condition Eq.~(\ref{p_quantize}) is applied. $N_{12}$ is the greatest common divisor of $N_1$ and $N_2$. Once $a_\mu^1$ and $a_\mu^2$ are turned on, the energetic cost of symmetry domain walls is compensated by gauge configurations. Thus, the above novel domain wall configurations proliferate leading to a symmetric ground state.

\section{Topological Quantum Field Theory of Symmetry-Protected Topological Phases}\label{sec:TQFTSPT}

 \subsection{Duality between GL theory and TQFT}
In the following, we may apply duality transformations to obtain a TQFT description of GL action.   We  start with the continuum action $S_0+S_{\rm Top}$ and derive TQFT.
Next step is to apply  {Hubbard-Stratonovich transformation}:
\begin{align}
 \mathcal{L}=& i \sum^2_I    n^I_\mu ( {a}^I_\mu-\partial_\mu\theta_s^I)
+\sum_I^2\frac{(f^I_{\mu\nu})^2}{4g^2}\,\nonumber\\
&-ip  \ep
  ( {a}_\mu^1-\partial_\mu \theta_s^1)   ( {a}_\nu^2-\partial_\nu\theta_s^2)   \partial_\lambda  {a}^2_{\rho} +\frac{1}{2\chi}\sum^2_I (n^I_\mu)^2\,,\nonumber
\end{align}
  where the vector fields $n_\mu^I$ are auxiliary fields in order to linearize the quadratic terms.
After total derivative terms are dropped off, integrating out $\theta_s^1$ leads to:
$ \partial_\mu (n^1_\mu-p {a}^2_\nu \partial_\lambda  {a}^2_\rho\ep )=0
 $
 which can be resolved by introducing a 2-form gauge field $b^1_{\mu\nu}$: ($b^{1}_{\mu\nu}=-b^{1}_{\nu\mu}$)
 \begin{align}
  n^1_\mu= p  {a}^2_\nu \partial_\lambda  {a}^2_\rho\ep+\frac{1}{4\pi}\partial_\nu b^1_{\lambda\rho}\ep\,.\nonumber\end{align}
 After replacing $n^1_\mu$ by the above identity, we end up with:
\begin{align}
\mathcal{L}=&i n^2_\mu ( {a}^2_\mu-\partial_\mu\theta_s^2)    + i\frac{1}{4\pi}  {a}^1_\mu \partial_\nu b^1_{\lambda\rho}\ep\nonumber\\
&+ip   \ep  {a}_\mu^1 \partial_\nu \theta_s^2  \partial_\lambda {a}^2_{\rho}+\sum^2_I\frac{(f^I_{\mu\nu})^2}{4g^2}+\sum^2_I \frac{1}{2\chi}(n^I_\mu)^2\,,\nonumber
 \end{align}
 where the last term will be replaced at the final step.

 Now, further integrating out $\theta_s^2$ leads to:
$\partial_\mu  (n^2_\mu+p  {a}^1_\nu \partial_\lambda  {a}^2_\rho\ep)=0
 $
 which can be resolved by introducing a 2-form gauge field $b^2_{\mu\nu}$: ($b^{2}_{\mu\nu}=-b^{2}_{\nu\mu}$)
 \begin{align}
  n^2_\mu=-p  {a}^1_\nu \partial_\lambda  {a}^2_\rho\ep+\frac{1}{4\pi}\partial_\nu b^2_{\lambda\rho}\ep\,.\nonumber
\end{align}
   After replacing $n^2_\mu$ by the above identity,   we end up with  the partition function with the TQFT action:
\begin{align}
S_{\rm TQFT}\!=\!\frac{i}{2\pi}\!\sum^2_I\! \int\! b^I\wedge d a^I\!\!+ip \int \!\!a^1\wedge a^2\wedge d a^2+S_{M},\!\!\label{a1a2da3_action}
\end{align}
where   $b^I_{\mu\nu}$ are three   2-form $\mathrm{U(1)}$ gauge fields.  As a side note, the differential form notation `$b^I$' is replaced by `$\frac{b^I_{\mu\nu}}{2!}$' once spacetime indices  are written explicitly.

    To the best of our knowledge,  the term  $\int a^1\wedge a^2\wedge da^2$ as a  classical action was  introduced in Ref.~\onlinecite{bod} where two additional constraints $da^1\wedge da^2=0$ and $da^2\wedge da^2=0$ are imposed to recover gauge invariance \cite{bod}.   And $S_{M}$, as an ultraviolet regulator, is given by:
\begin{align}S_{  M}=&\int d^4x\frac{1}{2\chi}(p  {a}^2_\nu \partial_\lambda  {a}^2_\rho\ep+\frac{1}{4\pi}\partial_\nu b^1_{\lambda\rho}\ep)^2\nonumber\\
&+\int d^4x\frac{1}{2\chi}(-p  {a}^1_\nu \partial_\lambda  {a}^2_\rho\ep+\frac{1}{4\pi}\partial_\nu b^2_{\lambda\rho}\ep)^2\nonumber\\
&+\sum_I^2\int d^4x\frac{(f^I_{\mu\nu})^2}{4g^2}.
\end{align}
Interestingly, from $S_M$, one can explicitly read the gauge transformations (\ref{gauge_b}) that play a very important role in the remaining discussions.

\subsection{Exotic gauge transformations}\label{sec:gauge}
 The first term in Eq.~(\ref{a1a2da3_action}) is two copies of topological BF term \cite{bf1,bf2,bf3} at level-1. As usual, all gauge fields are subject  to the following usual Dirac quantization conditions:
\begin{align}
 \frac{1}{2\pi}\int_{\mathcal{M}^2}  d{a}^I\in \mathbb{Z}\,\,,\,\,\, \frac{1}{2\pi } \int_{\mathcal{M}^3} db^I\in \mathbb{Z}\,, \label{constraint_aada}
\end{align}
where $\mathcal{M}^2$ is any closed two-dimensional surface.  As an Abelian gauge theory, the gauge transformations are given by:
 \begin{align}
a^I\rightarrow a^I+\,d\chi^I\,,\,b^I\rightarrow b^I+\,d V^I -2\pi p \,\epsilon^{IJ3} \chi^J \,d a^2\,,\label{gauge_b}
\end{align}
where $\chi^{I}\in\mathbb{R}$  and $V_{\mu}^{I}\in\mathbb{R}$  are   independent scalars and vectors, respectively.
Here, both 1-form $\,d\chi^I$ and 2-form $\,dV^I$ are closed but allowed to be non-exact if the following integers are nonzero on non-contractable manifolds.
 \begin{align}
&\int_{\mathcal{M}^1} \,d\chi^I=2\pi n^I\,,~~\int_{\mathcal{M}^2} \,dV^I=2\pi k^I\label{large_chi}\,,
 \end{align}
where $n^I$ and $k^I$ are six independent integers. For nonzero $n^I$ and $k^I$, the corresponding gauge transformations are said to be ``large gauge transformations'' labeled by the winding numbers $n^I$ and $k^I$.

\begin{figure}[t]
\centering
\includegraphics[width=7cm]{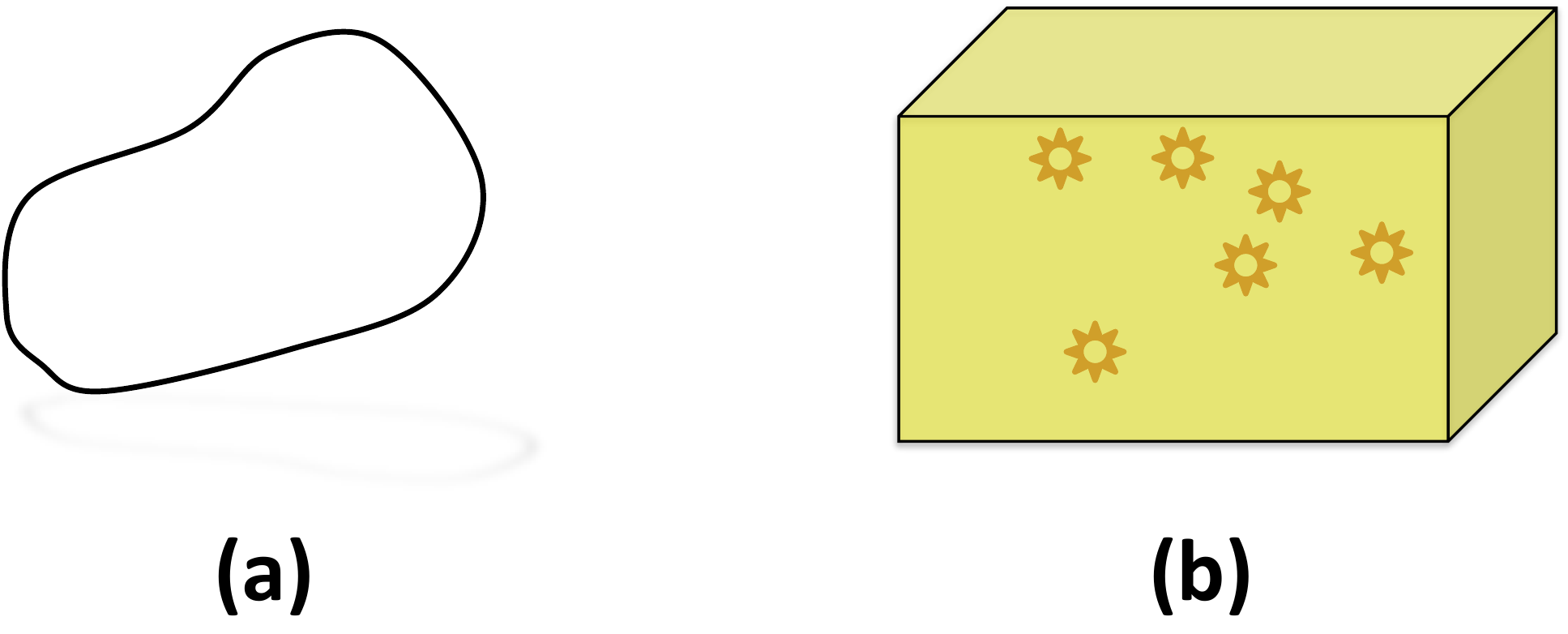}
\caption{{(Color online)} Illustration of gauge-invariant operators introduced in Sec.~\ref{sec:gauge}. (a) The Wilson loop operator $e^{i\int_{\mathcal{M}^1} a^I}$. (b) The operator formed by $\exp\{i\int_{\mathcal{M}^2} b^I-i2\pi p\int_{\mathcal{V}^3} \epsilon^{IJ3}a^J\wedge \,d a^2\}$ where $\mathcal{M}^2=\partial \mathcal{V}^3$. In (b), the cube represents $\mathcal{V}^3$ and its surface represents $\mathcal{M}^2$. The star symbols in (b) represent nonzero contributions of the ``Chern-Simons density'' $\int_{\mathcal{V}^3} \epsilon^{IJ3}a^J\wedge \,d a^2$ in $\mathcal{V}^3$.}
\label{figure_wilson}
\end{figure}

If $p=0$, $b^I$ are transformed in the usual definitions of 2-form gauge fields. Remarkably, the presence of $p$ term in Eq.~(\ref{a1a2da3_action}) induces a $p$-dependent term in the gauge transformations. Due to such unusual gauge transformations, we may construct the  following two gauge-invariant operators:
\begin{align}
& \text{Wilson loops:  }   e^{i\int_{\mathcal{M}^1} a^I}\nonumber\\
&\text{Modified Wilson surfaces:  } e^{i\int_{\mathcal{M}^2} b^I-i2\pi p\int_{\mathcal{V}^3} \epsilon^{IJ3}a^J\wedge \,d a^2}\nonumber
\end{align}
 where $\mathcal{V}^3$ is an open volume enclosed by $\mathcal{M}^2$, i.e., $\partial \mathcal{V}^3=\mathcal{M}^2$. It means that the Wilson surface operators of $b^1$ and $b^2$ are not gauge invariant alone, in contrast to the usual definitions. In Fig.~\ref{figure_wilson}, these gauge invariant operators are shown pictorially.

\subsection{Global symmetry $G= \mathbb{Z}_{N_1}\times \mathbb{Z}_{N_2}$ in TQFT}

In TQFT action (\ref{a1a2da3_action}), the physical meanings of the two form gauge field $b$ can be understood by identifying its curvature as four-current of point-particles: $J^I_\mu=\frac{1}{4\pi}\ep \partial_\nu b^I_{\lambda\rho}$. In terms of differential forms, we have:
\begin{align}
 {\star} J^I=\frac{1}{2\pi}  db^I\,,
\end{align}
where $\star$ denotes Hodge dual operation. If the symmetry is $\mathrm{U(1)}$, then, we may add a $\mathrm{U(1)}$ external background gauge potential  $A^I$ to impose the $\mathrm{U(1)}$ symmetry:
\begin{align}
S_{\rm TQFT}\rightarrow S=S_{\rm TQFT}+
\sum_I^2\frac{i}{2\pi}\int A^I
\wedge db^I\,,\label{StS}
\end{align}
where constant terms are neglected unless otherwise specified.
  In our case of $G=\prod_I^3\mathbb{Z}_{N_I}$, the particle current is conserved mod $N_I$.  To impose this condition, the closed loop integral of the 1-form  $A^I$ must be quantized at $\frac{2\pi}{N_I}$:
 \begin{align}
\int_{\mathcal{S}^1}A^I=\frac{2\pi}{N_I}\times 0,\pm 1,\cdots
 \end{align}
 \subsection{Symmetry-protected quantization}

 The implementation of the discrete symmetry leads to the following quantization and periodicity of $p$:
 \begin{align}
& p=\frac{k}{4\pi^2}\frac{N_1N_2}{N_{12}}\,,~~k\in\mathbb{Z}_{N_{12}}\,,\label{p_quantize}
 \end{align}
where $N_{12}$ denotes the greatest common divisor (GCD) of $N_1$ and $N_2$. Eq.~(\ref{p_quantize}) indicates that  there are totally ${N_{12}}$ topologically distinct  SPT  phases that are described by  TQFT Eq.~(\ref{a1a2da3_action}). For example, the  SPT  phase labeled by $k$ is equivalent to that labeled by $k+{N}_{12}$.
 In addition to Eq.~(\ref{a1a2da3_action}),  we may also consider a TQFT (as well as its related   Ginzburg-Landau   action) with $\frac{i}{2\pi}\sum^2_{I}\int b^I\wedge da^I+ip\int a^2\wedge a^1\wedge\,d a^1$, where $
 p=\frac{k}{4\pi^2}\frac{N_1N_2}{N_{12}}\,,~~k\in \mathbb{Z}_{N_{12}}$\,. As a result, the total classification of irreducible SPT phases with $\prod_I^2\mathbb{Z}_{N_I}$ symmetry is given by $(\mathbb{Z}_{N_{12}})^2$.

Physically, the quantization in Eq.~(\ref{p_quantize}) can be understood in the following way.   By performing the   gauge transformations (\ref{gauge_b}),  the total $ \mathbb{Z}_{N_I}$ symmetry charges ($I=1,2\,;\bar{I}=3-I$) are  not gauge invariant:
\begin{align}
\frac{1}{2\pi}\int_{\mathcal{M}^3} db^I\rightarrow \frac{1}{2\pi}\int_{\mathcal{M}^3} db^I-p\int_{\mathcal{M}^3} d\chi^{\bar{I}}\wedge da^2\,.\label{charge_change}
\end{align}
Since the 1-form external gauge potential $A^I$ is quantized at $\frac{2\pi}{N_I}$,  we may enforce the additional terms are divisible by $N_I$, i.e. $p\int_{\mathcal{M}^3} d\chi^{\bar{I}}\wedge da^2/N_I\in\mathbb{Z}$ such that the partition function is still invariant ($e^{iS}\rightarrow e^{iS+2\pi i}=e^{iS}$ with the action $S$ defined in Eq.~(\ref{StS})). Concretely speaking,
  $p$ should be quantized properly in order to ensure that there exists an integer pair $(m^1, m^2)$ such that the two equations below always hold: ($I=1,2$; $\bar{I}=3-I$.)
$  p\int_{\mathcal{M}^3}\,d\chi^{\bar{I}}\wedge \,da^2= N_I \, m_I\,
$.   To proceed further, let us consider $\mathcal{M}^3=S^1\times T^2$ such that for $I=1$: $
2\pi  p\int_{\mathcal{M}^3}\,d\chi^2\wedge \,da^2=2\pi p \int_{S^1}\,d\chi^2 \int_{T^2}\,da^2=2\pi p (2\pi)^2\times \text{integers}$
 where Eqs.~(\ref{constraint_aada},\ref{large_chi}) are applied.  Therefore, the existence of integer $m_1$ requires that ``$4\pi^2 p$ is divisable by $N_1$''. Likewise, we can obtain a similar condition:   ``$4\pi^2 p$ is divisable by $N_2$''. In summary, $p$ is quantized as:
$
 p=\frac{k}{4\pi^2}\frac{N_1N_2}{N_{12}}\,,~~k\in\mathbb{Z}\,,
 $ where $N_{12}$ is the greatest common divisor (GCD) of $N_1$ and $N_2$.

 However, $k\in\mathbb{Z}$ doesn't mean that the classification of bosonic SPT is  necessarily $\mathbb{Z}$. Let us consider  the following \textit{shift operation}:
  \begin{align}
  & d b^1\rightarrow d b^1-\frac{K^1 N_1N_2}{2\pi N_{12}}a^2\wedge d a^2\,\,,\label{shift1}
  \\
  & d b^2\rightarrow d b^2+\frac{K^2 N_1N_2}{2\pi N_{12}}a^1\wedge  d a^2\,\,,\label{shift2}
  \\
&k\rightarrow k+K^1+K^2\label{shift_of_smallk}\,,
  \end{align}
where $K^1$ and $K^2$ are two integers parametrizing the above shift operation.
Likewise, one can check that TQFT partition function is  invariant  under the above shift operation if and only if
  \begin{align}
 &\frac{1}{2\pi} \int_{\mathcal{M}^3}d b^{I}\bigg|_{\text{after shift}}-  \frac{1}{2\pi} \int_{\mathcal{M}^3}d b^I\bigg|_{\text{before shift}}\nonumber\\
& \text{is divisible by} \,N_I   \,\,\,(\text{  for $I=1,2$})\,.\label{definition_shift}
  \end{align}
 As a result, $e^{iS}\rightarrow e^{iS+2\pi i}=e^{iS}$ with the action $S$ defined in Eq.~(\ref{StS}).
By using the shift operations Eqs.~(\ref{shift1},\ref{shift2}), the two equations in Eq.~(\ref{definition_shift})  reduce to:
  \begin{align}
 &\frac{K^1N_1N_2}{4\pi^2 N_{12}} \int a^2 \wedge da^2 \text{ is divisible by }N_1 \,,\nonumber\\
 &\frac{K^2N_1N_2}{4\pi^2 N_{12}} \int a^1 \wedge da^2 \text{ is divisible by }N_2.\nonumber
\end{align}
  Integrating out $b^I$ leads to: $\int_{S^1}a^I=-\int_{S^1}A^I=\frac{2\pi}{N_I}\times 0,\pm 1, \cdots.$
   By  further using  the Dirac conditions (\ref{constraint_aada}), we end up with two conditions:
 $\frac{K^1N_1N_2}{4\pi^2 N_{12}}\frac{4\pi^2}{N_2}$   is divisible by $N_1$ and $\frac{K^2N_1N_2}{4\pi^2 N_{12}}   \frac{4\pi^2}{N_1}$  is divisible by $N_2$.    The general solution is $K^1=l N_{12}, K^2=l'N_{12}$ with $l,l'\in\mathbb{Z}$. Therefore, by using Bezout's lemma, $K^1+K^2$ is quantized  at  GCD of  $N_{12}$ and $N_{12}$, i.e., $N_{12}$. In other words, the minimal shift of $k$ [see Eq.~(\ref{shift_of_smallk})] is  $N_{12}$.
For our bosonic SPT phases, let us start with $k=0$ which describes the trivial phase. Then, by increasing $k$, distinct bosonic SPT phases labeled by $k=1,2,\cdots$ are obtained in succession. But once $k=N_{12}$, the state returns to the trivial state immediately since it can be connected to $k=0$ trivial state just by the above   shift operation. One may also consider TQFT with action:
 \begin{align}
S_{\rm TQFT}\!=\!\frac{i}{2\pi}\!\sum^2_I\! \int\! b^I\!\wedge \!d a^I\!\!+ip \int \!\!a^2\wedge \!a^1\wedge d a^1+\!S_{M} \!\!\label{a1a2da3_actionAA1}
\end{align}
 which leads to another set of  SPT  states with $p=\frac{kN_1N_2}{4\pi^2 N_{12}}$ and classified by $k\in \mathbb{Z}_{N_{12}}$. As a result, the total number of distinct 3D  SPT  states with $\prod_I^2\mathbb{Z}_{N_I}$  symmetry is given by the cyclic group $(\mathbb{Z}_{N_{12}})^2$, which is also consistent to group cohomology: $\mathcal{H}^{4}(\prod^2_I\mathbb{Z}_{N_I},\mathrm{U(1)})  =(\mathbb{Z}_{N_{12}})^2$.

\section{Other symmetries and applications}\label{sec:general_symmetry}

\subsection{Other Abelian symmetries}

 In Appendix \ref{appendix_123}, the \textit{irreducible} (see Table \ref{table1}) SPT phases with $\mathbb{Z}_{N_1}\times\mathbb{Z}_{N_2}\times\mathbb{Z}_{N_3}$ are derived, where the actions of TQFT are  given by:
 \begin{align}
S_{\rm TQFT}\!=\!\frac{i}{2\pi}\!\sum^3_I\! \int\! b^I\wedge d a^I\!\!+ip \int \!\!a^1\wedge a^2\wedge d a^3+S_{M},\!\!\label{a1a2da3_actionAA}
\end{align}
and,
\begin{align}
S_{\rm TQFT}\!=\!\frac{i}{2\pi}\!\sum^3_I\! \int\! b^I\wedge d a^I\!\!+ip \int \!\!a^2\wedge a^3\wedge d a^1+S_{M}.\!\!\label{a1a2da3_actionAAA}
\end{align}
(\ref{a1a2da3_actionAA}) and (\ref{a1a2da3_actionAAA}) produce $(\mathbb{Z}_{N_{23}})^2$ classification of irreducible SPT phases.
Since   SPT  phases with only two $ \mathbb{Z}_N$ symmetries can be viewed as \textit{reducible}  SPT  phases with three $ \mathbb{Z}_N$ symmetries, we conclude that  from TQFT approach the full classification of  SPT  phases with G$=\prod^3_I\mathbb{Z}_{N_I}$ symmetry is given by:
$\mathcal{H}^{4}(\prod^3_I \mathbb{Z}_{N_I},\mathrm{U(1)})  =(\mathbb{Z}_{N_{12}})^2\times(\mathbb{Z}_{N_{23}})^2\times(\mathbb{Z}_{N_{13}})^2   \times (\mathbb{Z}_{N_{123}})^2\,. $  Therefore, we have achieved all  SPT  phases with either $\prod^3_{I} \mathbb{Z}_{N_I}$ or $\prod^2_{I} \mathbb{Z}_{N_I}$ symmetries that were predicted in group cohomology.

 The irreducible SPT phases with  $\prod^4_{I}\mathbb{Z}_{N_I}$ symmetry have been studied in Ref.~\onlinecite{GWW} where TQFT is given by:
\begin{align}
S_{\rm TQFT} =\frac{i}{2\pi}\sum^4_I \int \!\! b^I\wedge \!d a^I+ip \int \!\!a^1\!\!\wedge \!a^2\!\!\wedge\! a^3\wedge  \!a^4\,
\end{align} with   conditions Eq.~(\ref{constraint_aada})   with $I=1,\cdots,4$.  $p=\frac{kN_1N_2N_3N_4}{(2\pi)^3N_{1234}}$ with $k\in\mathbb{Z}_{N_{1234}}$ where $N_{1234}$ is GCD of all $N_I$'s. Thus, for arbitrary finite unitary Abelian group $G=\mathbb{Z}_{N_1}\times \mathbb{Z}_{N_2}\cdots$, we can derive the TQFT descriptions for all SPT phases classified by the group cohomology:
\begin{align}
&\mathcal{H}^{4}(\mathbb{Z}_{N_1}\times \mathbb{Z}_{N_2}\cdots,\mathrm{U(1)})\nonumber\\
=&\Pi_{I<J}(\mathbb{Z}_{N_{IJ}})^2\!\times\!\Pi_{I<J<K}(\mathbb{Z}_{N_{IJK}})^2\!\times\!\Pi_{I<J<K<L}\mathbb{Z}_{N_{IJKL}}\nonumber.
\end{align} In fact, all the results for finite unitary Abelian group can be generalized into cases that  involve $\mathrm{U(1)}$ symmetry as well. In Appendices, we show that there is no nontrivial SPT with $\mathrm{U(1)}^k$ or $\mathbb{Z}_{N_1}\times \mathrm{U(1)}^k$ symmetry, and the irreducible SPT phase with $\prod^2_I\mathbb{Z}_{N_I}\times \mathrm{U}(1)$ symmetry can be described by  $\frac{i}{2\pi}\int \sum^3_Ib^I\wedge \,da^I+ip\int  a^1\wedge a^2\wedge \,da^3$, where $p=\frac{k}{4\pi^2}\frac{N_1N_2}{N_{12}}\,,~k\in\mathbb{Z}_{N_{12}}\,$ gives rise to a $\mathbb{Z}_{N_{12}}$ classification. Together with the $(\mathbb{Z}_{N_{12}})^2$ reducible SPT phases contributed from SPT phases protected merely by the $\prod_I^2\mathbb{Z}_{N_I}$ part, we derive the TQFT descriptions for all the $(\mathbb{Z}_{N_{12}})^3$ SPT phases predicted by group cohomology classifications. The above scheme can be easily generalized to arbitrary Abelian group symmetry with the form $ \tilde G=\mathbb{Z}_{N_1}\times \mathbb{Z}_{N_2}\cdots\times \mathrm{U(1)}^k \equiv G\times \mathrm{U(1)}^k$, that is, in addition to the TQFT descriptions for finite unitary Abelian group piece $G$, one can always pick up a subgroup $\mathbb{Z}_{N_I}\times \mathbb{Z}_{N_J} \times\mathrm{U(1)}$ to construct $\mathbb{Z}_{N_{IJ}}$ additional SPT phases.


 \subsection{ {{Connection to  ``decorated domain walls'' picture }}}
Ref.~\onlinecite{2d19} proposed a way called ``decorated domain walls'' to construct the ground state wave function of  SPT  states. 
   Take $\prod^2_I\mathbb{Z}_{N_I}$ symmetry as an example, the  SPT  ground state can be formed as an equal weight superposition of 2D domain walls of $\mathbb{Z}_{N_1}$ on which a nontrivial 2D $ \mathbb{Z}_{N_2}$  SPT  is placed. Here we use our TQFT approach to understand the ``decorated domain walls'' picture. Let us start with  the action $S$ in Eq.~(\ref{StS}) and set $A^1=0, A^2=A$. We also add a dynamical Higgs term:
  \begin{align}
   {S}_{\rm Higgs}= \int\!\frac{iN_I}{2\pi}B\wedge d A\,,
 \end{align}
 where $A$ and $B$ are 1-form and 2-form dynamical gauge fields respectively. $ {S}_{\rm Higgs}$ breaks $\mathrm{U(1)}$ gauge symmetry of $A^2$ gauge field down to $ \mathbb{Z}_{N_2}$.
 Integrating out $B$ and $b^2$ leads to $\int_{\mathcal{M}^1}a^2=\int_{\mathcal{M}^1}A=-\frac{2\pi}{N_2}\times\text{integer}$. $a^1\wedge a^2\wedge da^2$ becomes: $
ip\int a^1\wedge A \wedge \,dA$. $a^1$ can be viewed as domain wall of $\mathbb{Z}_{N_1}$ symmetry such that it may be replaced by $\frac{2\pi  k_1}{N_1}$ where $  k_1\in\mathbb{Z}_{N_1}$. In other words, given a  $\mathbb{Z}_{N_1}$ symmetry domain wall ``$\mathcal{D}(\mathbb{Z}_{N_1})$'' labeled by $k_1$, there is a topological gauge theory on this 2D manifold:
\begin{align}
S[A]=&\frac{i}{4\pi}\int_{\mathcal{D}(\mathbb{Z}_{N_1})}\!\!  \frac{2N_2k_1k}{N_{12}}A\wedge d A\,,\label{topological_response_action}
\end{align}
where  we have used $p=\frac{k}{4\pi^2}\frac{N_1N_2}{N_{12}}$ with $k\in\mathbb{Z}_{N_{12}}$.
 By noting that  $\frac{2N_2k_1k}{N_{12}}$ is even\cite{2d8,2d1,2d_pollmann,2d5,2d6, 2d11,2d10},   Eq.~(\ref{topological_response_action}) suggests that the 2D state decorated on the domain wall is indeed a nontrivial 2D  SPT  with $\mathbb{Z}_{N_2}$ symmetry.
 Such a way of thinking can be naturally extended to   SPT  phases with $\prod^3_I\mathbb{Z}_{N_I}$   described by TQFT Eq.~(\ref{a1a2da3_action}). For this purpose, we may introduce $A$ and $\tilde{A}$ that respectively gauge $\mathbb{Z}_{N_2}$ and $\mathbb{Z}_{N_3}$ symmetry groups. As a result, on the $\mathbb{Z}_{N_1}$ symmetry domain wall, we have the topological gauge action for $\mathbb{Z}_{N_2}\times \mathbb{Z}_{N_3}$  SPT  phases:
 \begin{align}
 S[A,\tilde{A}]=\frac{i}{4\pi}\int_{\mathcal{D}(\mathbb{Z}_{N_1})}  \frac{2N_2k_1k}{N_{12}}A\wedge \,d \tilde{A}\,,
\end{align}
 where  Eq.~(\ref{p_quantize}) is applied and $k\in\mathbb{Z}_{N_{123}}$.

\subsection{ {{Connection to 3-loop statistics}}}
  It was first shown in Ref.~\onlinecite{wang_levin1} that, after the global symmetry $\mathbb{Z}_{N_1}\times\mathbb{Z}_{N_2}\times\cdots$ of  SPT  phases is gauged at weak gauge coupling limit, the loop excitations of resultant gapped state exhibit the so-called 3-loop statistics.
   Based on our TQFT approach to  SPT with $G=\mathbb{Z}_{N_1}\times\mathbb{Z}_{N_2}$ , we may formally gauge the symmetry by adding  the coupling term:
\begin{align}
    {S}_{\rm coupling}\!= \!\int\!\sum^2_I\!\!\frac{i}{2\pi}A^I\wedge d b^I+\int\!\sum_I^2\!\frac{iN_I}{2\pi}B^I\!\wedge \!d A^I\,,
 \end{align}
 where $A^I$ and $B^I$ are 1-form and 2-form dynamical gauge fields respectively. The first term is the minimal coupling term between $A^I$ and matter field current $\star\frac{1}{2\pi}d b^I$; the second term is a Higgs term that breaks $\mathrm{U(1)}$ gauge symmetry of $A^I$ gauge field down to $\mathbb{Z}_{N_I}$. After integrating out $a^I$ and $b^I$, it turns out that the resultant new TQFT is given by (e.g. symmetry G$=\prod^3_I\mathbb{Z}_{N_I}$):
  \begin{align}
S&=\sum^2_I \int \frac{iN_I}{2\pi}B^I\wedge \,d A^I+ip \int A^1\wedge A^2\wedge \,d A^2\,.\label{gauge_a1a2da2}
\end{align}
 By integrating out   $B^I$, we will end up with an action for Dijkgraaf-Witten \cite{dwitten} gauge theory proposed in Ref.~\onlinecite{corbodism3}.
 We conjecture that Eq.~(\ref{gauge_a1a2da2})   gives rise to the   3-loop statistics. Some progress has been made along this line  \cite{3loop_ryu,cjw_gsd,levin_talk}.

 \section{Conclusions and future directions}\label{sec:conclusions}
  In this paper, we attempt to provide a complete TQFT description for all 3D SPT phases with unitary Abelian group symmetry. Taking $\prod_I^2\mathbb{Z}_{N_I}$ as an example, we start with a Ginzburg-Landau type action and illustrate the key mechanism via  proliferating ``nontrivial" symmetry domain walls for SPT phases with finite unitary Abelian group symmetry. We then rigorously derive the corresponding TQFT  and compute the level quantization. Finally, we consider generic 3D SPT phases with arbitrary unitary Abelian group symmetry, including $\mathrm{U(1)}$. All irreducible TQFT results for unitary Abelian group symmetry are collected in Table \ref{table1}.
In Appendices, we also discuss possible TQFT descriptions with anti-unitary time reversal symmetry.
Together with the previous work of bosonic topological insulators \cite{yegu2014}, we provide a route towards a complete TQFT description for 3D SPT phases and reveal the key mechanism of these exotic quantum phases of matter. 

There are several interesting directions for future studies. One of them is to construct  the boundary of 3D SPT phases by using the present TQFT framework. There are two steps. First, we should consider TQFT in an open manifold and try to derive the boundary effective field theory. It is possible that the gauge invariant argument in boundary CFT derivation of Chern-Simons theory \cite{Wenbook} is still applicable in the present TQFT. However, the exotic gauge transformations, e.g. Eq.~(\ref{gauge_b}), may lead to technical challenge.  Second, the symmetry transformations on the boundary effective field theory should reveal some nontrivial properties that forbid the realization of the boundary theory in a 2D plane alone (i.e. no 3D bulk).
In addition, it is also interesting to study the symmetry-enriched topological phases (SET) by using TQFT. SET has  topological excitations in the bulk and symmetry acts on excitations in a fractionalized manner (e.g. projective representation). For this purpose, one may replace $\frac{1}{2\pi}\sum_Ib^I\wedge da^I$ by $\sum_I\frac{k_I}{2\pi}b^I\wedge da^I$ with $k_I=2,3,\cdots$ and then study how to impose symmetry $G$. Along this line, one may study  3D SET  phases with gauge group $\mathbb{Z}_{k_1}\times\mathbb{Z}_{k_2}\times\cdots$ and symmetry group $G$.
Finally, it is interesting to extend all of the above studies   to fermionic SPT and fermionic SET where the transparent quasiparticles are fermionic (e.g. electrons in FQHE).

 \section*{Acknowledgement}
  We would like to thank Q.-R. Wang,  Ken Shiozaki, AtMa Chan, A. Tiwari, M. Stone, Xiao Chen,   E. Fradkin, T. Hughes, S. Ryu,  D. Gaiotto, X.-G. Wen, A. Kapustin, N. Seiberg and S.-T. Yau's discussions. Research at Perimeter Institute is supported by the Government of Canada through Industry Canada and by the Province of Ontario through the Ministry of Economic Development \& Innovation.(P.Y. and Z.C.G.)  P.Y. is supported in part by  the NSF through grant DMR 1408713 at the University of Illinois.  We also acknowledge the warm hospitality and support from the Center of Mathematical Sciences and Applications at Harvard University where the work was done in part.
\appendix
\section{TQFT of irreducible bosonic SPT phases with  global symmetry $\mathbb{Z}_{N_1}\times\mathbb{Z}_{N_2}\times\mathbb{Z}_{N_3}$}\label{appendix_123}

In this Appendix, we derive TQFT and its classification when the symmetry is  $\mathbb{Z}_{N_1}\times\mathbb{Z}_{N_2}\times \mathbb{Z}_{N_3}$. The action is given by:
\begin{align}
S_{\rm TQFT}\!=\!\frac{i}{2\pi}\!\sum^3_I\! \int\! b^I\wedge d a^I\!\!+ip \int \!\!a^1\wedge a^2\wedge d a^3+S_{M}\!\!\label{a1a2da3_actionA}
\end{align}
 The implementation of the discrete symmetry leads to the following quantization and periodicity of $p$:
 \begin{align}
& p=\frac{k}{4\pi^2}\frac{N_1N_2}{N_{12}}\,,~~k\in\mathbb{Z}_{N_{123}}\,,\label{p_quantizeA}
 \end{align}
where $N_{12}$ ($N_{123}$) denotes the greatest common divisor (GCD) of $N_1$ and $N_2$ ($N_1$, $N_2$ and $N_3$). Eq.~(\ref{p_quantizeA}) indicates that  there are totally ${N_{123}}$ topologically distinct  SPT  phases that are described by  TQFT Eq.~(\ref{a1a2da3_actionA}). For example, the  SPT  phase labeled by $k$ is equivalent to that labeled by $k+{N}_{123}$.
 In addition to Eq.~(\ref{a1a2da3_actionA}),  we may also consider a TQFT (as well as its related   Ginzburg-Landau   action) with $\frac{i}{2\pi}\sum^3_{I}\int b^I\wedge da^I+ip\int a^2\wedge a^3\wedge\,d a^1$, where $
 p=\frac{k}{4\pi^2}\frac{N_3N_2}{N_{23}}\,,~~k\in \mathbb{Z}_{N_{123}}$\,. As a result, the total classification of irreducible SPT phases with $\prod_I^3\mathbb{Z}_{N_I}$ symmetry is given by $(\mathbb{Z}_{N_{123}})^2$.

Physically, the quantization in Eq.~(\ref{p_quantizeA}) can be understood in the following way.   By performing the   gauge transformations (\ref{gauge_b}),  the total $\mathbb{Z}_{N_I}$ symmetry charges ($I=1,2\,;\bar{I}=3-I$) are  not gauge invariant:
\begin{align}
\frac{1}{2\pi}\int_{\mathcal{M}^3} db^I\rightarrow \frac{1}{2\pi}\int_{\mathcal{M}^3} db^I-p\int_{\mathcal{M}^3} d\chi^{\bar{I}}\wedge da^3\,.\label{charge_changeA}
\end{align}
Since the 1-form external gauge potential $A^I$ is quantized at $\frac{2\pi}{N_I}$,  we may enforce the additional terms are $N_I\times $integer: $p\int_{\mathcal{M}^3} d\chi^{\bar{I}}\wedge da^3=N_I\times\mathbb{Z}$ such that the partition function is still invariant. Concretely speaking,
  $p$ should be quantized properly in order to ensure that there exists an integer pair $(m^1, m^2)$ such that the two equations below always hold: ($I=1,2$; $\bar{I}=3-I$.)
$  p\int_{\mathcal{M}^3}\,d\chi^{\bar{I}}\wedge \,da^3= N_I \, m_I\,$.   To proceed further, let us consider $\mathcal{M}^3=S^1\times T^2$ such that for $I=1$:
\begin{align}
2\pi  p\int_{\mathcal{M}^3}\,d\chi^2\wedge \,da^3=&2\pi p \int_{S^1}\,d\chi^2 \int_{T^2}\,da^3
\nonumber\\
=&2\pi p (2\pi)^2\times \text{integers}\,,\nonumber
 \end{align}
 where Eqs.~(\ref{constraint_aada},\ref{large_chi}) are applied.  Therefore, the existence of integer $m_1$ requires that ``$4\pi^2 p$ is divisable by $N_1$'' such that $m_1=\frac{4\pi^2 p}{N_1}\times \text{integer}\in \mathbb{Z}$. Likewise, we can obtain a similar condition:   ``$4\pi^2 p$ is divisable by $N_2$''. In summary, $p$ is quantized as:
$
 p=\frac{k}{4\pi^2}\frac{N_1N_2}{N_{12}}\,,~~k\in\mathbb{Z}\,,
 $ where $N_{12}$ is the greatest common divisor (GCD) of $N_1$ and $N_2$.

 However, $k\in\mathbb{Z}$ doesn't mean that the classification of bosonic SPT is  necessarily $\mathbb{Z}$. Let us consider  the following \textit{shift operation}:
  \begin{align}
&  b^3\rightarrow b^3-\frac{K^3 N_1N_2}{2\pi N_{12}}a^1\wedge a^2\,\,,\label{shift3A}
 \\
  & d b^1\rightarrow  d b^1-\frac{K^1 N_1N_2}{2\pi N_{12}}a^2\wedge da^3\,\,,\label{shift1A}
  \\
  & db^2\rightarrow d b^2-\frac{K^2 N_1N_2}{2\pi N_{12}}da^3\wedge a^1\,\,,\label{shift2A}
  \\
&k\rightarrow k+K^1+K^2+K^3\label{shift_of_k}\,,
  \end{align}
where $K^1$, $K^2$, and $K^3$ are three integers parametrizing the above shift operation. 
 One can check that TQFT is formally invariant  under the above shift operation.   Furthermore, the shifts in $b^I$ fields should be consistent to the $\mathbb{Z}_{N_I}$ symmetry  via the following relations:
  \begin{align}
 &\frac{1}{2\pi} \int_{\mathcal{M}^3}d b^{I}\bigg|_{\text{after shift}}-  \frac{1}{2\pi} \int_{\mathcal{M}^3}d b^I\bigg|_{\text{before shift}}\nonumber\\
 = &\,N_I \times \text{integer} \,\,\,(\text{  for $I=1,2,3$})\label{definition_shiftA}
  \end{align}
  which means that the change amount should be divisible by $N_I$ for index $I$.
By using the shift operations Eqs.~(\ref{shift3A},\ref{shift1A},\ref{shift2A}), these relations  reduce to:
  \begin{align}
 &\frac{K^3N_1N_2}{4\pi^2 N_{12}} \left(\int a^1 \wedge da^2+ \int da^1 \wedge a^2\right)\text{ is divisible by }N_3\,,\nonumber\\
 &\frac{K^1N_1N_2}{4\pi^2 N_{12}} \int a^2 \wedge da^3 \text{ is divisible by }N_1\,,\nonumber\\
 &\frac{K^2N_1N_2}{4\pi^2 N_{12}} \int da^3 \wedge a^1\text{ is divisible by }N_2.\nonumber
\end{align}
It means the integers ($K^1,K^2,K^3$) should be properly selected such that the three ``integers'' on the right hand sides always exist. Integrating out $b^I$ leads to:
\begin{align}
\int_{S^1}a^I=-\int_{S^1}A^I=\frac{2\pi}{N_I}\times 0,\pm 1, \cdots.\label{aaa12345A}
\end{align}
 By   using Eqs.~(\ref{constraint_aada},\ref{aaa12345A}), the above equations are reexpressed as:
\begin{align}
 &\frac{K^3N_1N_2}{4\pi^2 N_{12}} \frac{4\pi^2\left( N_1\times\text{integer}+N_2\times\text{integer}\right)}{N_1N_2}\nonumber\\
 &\text{ is divisible by }N_3\,,\label{2pK1A}\\
 &\frac{K^1N_1N_2}{4\pi^2 N_{12}} \frac{4\pi^2}{N_2}
 \text{ is divisible by }N_1\,,\label{2pK3A}\\
 &\frac{K^1N_1N_2}{4\pi^2 N_{12}} \frac{4\pi^2}{N_1}
\text{ is divisible by }N_2\,.\label{2pK2A}
\end{align}
By noting that $\left( N_1\times\text{integer}+N_2\times\text{integer}\right)$ is always divisible by $N_{12}$ due to B\'{e}zout's lemma, we obtain the following solution:
$K^3/N_{3} \in\mathbb{Z}$, $K^1/N_{12}\in\mathbb{Z}$, and $K^2/N_{12}\in\mathbb{Z}$. By using B\'{e}zout's lemma again, we obtain the minimal shift of $k$ is $GCD(N_{12}, N_3)=GCD(N_1,N_2,N_3)\equiv N_{123}$.  Thus, for our bosonic SPT phases, let us start with $k=0$ which is trivial. Then, by increasing $k$, distinct bosonic SPT phases labeled by $k=1,2,\cdots$ are obtained in succession. But once $k=N_{123}$, the state returns the trivial state immediately since it can be connected to $k=0$ trivial state just by the above   shift operation.

\section{TQFT of irreducible bosonic SPT phases with  global symmetry $(\mathbb{Z}_{N_1}\times\cdots)\times \mathrm{U}(1)$}\label{appendix_u1u1}

\subsection{$\mathbb{Z}_N\times\mathrm{U(1)}$}

In the following, we consider bosonic SPT phases with direct product of $\mathrm{U(1)}$ and several $\mathbb{Z}_N$'s. First, let us consider
G$=\mathbb{Z}_{N}  \times \mathrm{U}(1)$. In this case,  we may consider TQFT  with the conditions:
\begin{align}
& \text{Dirac conditions: }\frac{1}{2\pi}\int_{\mathcal{M}^2}  d\,{a}^I\in \mathbb{Z}\,,\, \frac{1}{2\pi } \int_{\mathcal{M}^3} \,db^I\in\mathbb{Z}
 \\
& \text{Symmetry: }\,\frac{1}{2\pi } \int_{\mathcal{M}^3} \,db^1\rightarrow \frac{1}{2\pi } \int_{\mathcal{M}^3} \,db^1+N
\label{quantization_bfaada33}
\end{align}
which means that all bosons of $I=1$ carry $\mathbb{Z}_{N}$ symmetry while all bosons of $I=2$ carry $\mathrm{U(1)}$ symmetry. The gauge transformations for both $b^I$ are given by:
$  b^I\rightarrow    b^I -2\pi p \,\epsilon^{IJ} \chi^J  \,d a^2\,.
$ Again, we require that  the additional terms in $b^I$ fields  after gauge transformations can be removed by  the $\mathbb{Z}_N$ symmetry transformations Eq.~(\ref{quantization_bfaada33}). However, this can only be done for $I=1$. The absence of $I=2$ in Eq.~(\ref{quantization_bfaada33}) leads to $p=0$.

In a similar way, we may  show that $bda+a^2a^1da^1$ is also trivial. Note that bosonic SPT phases with either $\mathrm{U(1)}$ or $\mathbb{Z}_N$ symmetry are always trivial, we conclude that all bosonic SPT phases (both irreducible and reducible) with  G$=\mathbb{Z}_{N} \times \mathrm{U}(1)$ symmetry are trivial.

\subsection{$\prod_I^2\mathbb{Z}_{N_I}\times \mathrm{U}(1)$}
 Next, we consider  G$=\prod_I^2\mathbb{Z}_{N_I}\times \mathrm{U}(1)$. Since there are three independent symmetry groups, we need to consider TQFT in Eq.~(\ref{a1a2da3_action}). If we assume that the bosons of either $I=1$ or $I=2$ carry $\mathrm{U(1)}$ symmetry, then, it is still concluded that no irreducible bosonic SPT phases exist. Therefore, it is sufficient to only explore the possibility of nontrivial irreducible bosonic SPT phases where the bosons of $I=3$ carry $\mathrm{U(1)}$ symmetry. Then,  the following conditions should be imposed:
\begin{align}
&\text{Dirac conditions: }\frac{1}{2\pi}\int_{\mathcal{M}^2}  d{a}^I\in \mathbb{Z}\,,\,\frac{1}{2\pi } \int_{\mathcal{M}^3} \,db^I\in \mathbb{Z};
 \\
& \text{Symmetry: } \,\frac{1}{2\pi } \int_{\mathcal{M}^3} \,db^{I}\rightarrow \frac{1}{2\pi } \int_{\mathcal{M}^3} \,db^I+ N_I\nonumber\\
&~~~(I=1,2)\,.
\label{quantization_bfaada55}
\end{align}
Likewise, we can obtain the quantization of $p$: $ p=\frac{k}{4\pi^2}\frac{N_1N_2}{N_{12}}\,,~~k\in\mathbb{Z}$. We must also proceed further with the shift operation defined in  Eqs.~(\ref{shift3A},\ref{shift1A},\ref{shift2A},\ref{shift_of_k}). By taking account of the symmetry transformations Eq.~(\ref{quantization_bfaada55}),  Eq.~(\ref{definition_shiftA}) is changed to:
  \begin{align}
 &\frac{1}{2\pi} \int_{\mathcal{M}^3}d b^{I}\bigg|_{\text{after shift}}-  \frac{1}{2\pi} \int_{\mathcal{M}^3}d b^I\bigg|_{\text{before shift}}\nonumber\\
 = &\,N_I \times \text{integer} (\text{  for I=1,2})\label{definition_shift111}\\
 &\frac{1}{2\pi} \int_{\mathcal{M}^3}d b^{3}\bigg|_{\text{after shift}}-  \frac{1}{2\pi} \int_{\mathcal{M}^3}d b^3\bigg|_{\text{before shift}}=0 \label{definition_shift33}
  \end{align}
 Eq.~(\ref{definition_shift33})  leads to:
\begin{align}
K^3=0\,.\label{kkk222}
\end{align}
By using the shift operations Eqs.~(\ref{shift1A},\ref{shift2A}), Eq.~(\ref{definition_shift111}) leads to
  \begin{align}
 &\frac{K^1N_1N_2}{4\pi^2 N_{12}} \int a^2 \wedge da^3 \text{ is divisible by }N_1\,,\nonumber\\
 &\frac{K^2N_1N_2}{4\pi^2 N_{12}} \int da^3 \wedge a^1\text{ is divisible by }N_2.\nonumber
\end{align}
   By further using Eq.~(\ref{aaa12345A}) ($I=1,2$),    we end up with:
\begin{align}
 &\frac{K^1N_1N_2}{4\pi^2 N_{12}} \frac{4\pi^2}{N_2}
 \text{ is divisible by }N_1\,,\nonumber\\
 &\frac{K^1N_1N_2}{4\pi^2 N_{12}} \frac{4\pi^2}{N_1}
\text{ is divisible by }N_2\,,\nonumber
\end{align}
which means that: $K^1/N_{12}\in\mathbb{Z}\,,K^2/N_{12}\in\mathbb{Z}$.
Thus the minimal shift of $k$ is $N_{12}$.  Therefore, the irreducible  bosonic SPT  phases are classified by cyclic group $\mathbb{Z}_{N_{12}}$.

In summary, the complete classification of  bosonic SPT  phases with $\prod_I^2\mathbb{Z}_{N_I}\times \mathrm{U}(1)$ symmetry is given by $\mathbb{Z}_{N_{12}}\times (\mathbb{Z}_{N_{12}})^2=(\mathbb{Z}_{N_{12}})^3$ where the additional two $\mathbb{Z}_{N_{12}}$ parts arise from  bosonic SPT  phases with $\prod_I^2\mathbb{Z}_{N_I}$ symmetry only.

\section{TQFT of irreducible  bosonic SPT  phases with   time-reversal symmetry ($\mathbb{Z}_2^T$)}\label{appendix_tmtm}
\subsection{$\mathbb{Z}_2^T$ and $\mathrm{U(1)}\rtimes \mathbb{Z}_2^T$}\label{appendix_time_reversal}
The complete classification of  bosonic SPT  phases with   $\mathbb{Z}_2^T$ (time-reversal symmetry with $\mathcal{T}^2=1$) is given by $(\mathbb{Z}_2)^2$  \cite{Wencoho}. One $\mathbb{Z}_2$ index corresponds to  bosonic SPT  phases with surface ``all-fermion topological order'' \cite{2d61}. Its TQFT description is proposed to be the form of  ``$b\wedge da +b\wedge b$'' in Ref.~\onlinecite{yegu2014}. The other $\mathbb{Z}_2$ index corresponds to  bosonic SPT  phases with surface $\mathbb{Z}_2$ topological order where both $e$ and $m$ carry Kramers' doublet. Its TQFT description is just a multi-component $b\wedge d a$ theory of level-1 but with unusual definition of bulk time-reversal symmetry transformation, as shown in Section  VII of Ref.~\onlinecite{yegu2014}. These  bosonic SPT  can be viewed as either bosonic topological insulators where $\mathrm{U(1)}$ symmetry doesn't play role of symmetry protection or ``bosonic topological superconductors'' where $\mathrm{U(1)}$ symmetry is completely broken.

 Formally, $\mathrm{U(1)}\rtimes$$\mathbb{Z}_2^T$  is a \emph{non-Abelian} symmetry group due to the semi-product operation ``$\rtimes$'' but we still  summarize known results below. The complete classification of  bosonic SPT  phases with   U(1)$\rtimes$$\mathbb{Z}_2^T$ symmetry is given by $(\mathbb{Z}_2)^3$ \cite{Wencoho},  among which there are two $\mathbb{Z}_2$ indices are given by    bosonic SPT  phases with merely time-reversal symmetry. The remaining one $\mathbb{Z}_2$ index  labels irreducible  bosonic SPT  phases  with $\mathrm{U(1)}\times$$\mathbb{Z}_2^T$ symmetry so that both $\mathrm{U(1)}$ and $\mathbb{Z}^T_2$ play nontrivial role of symmetry protection. In other words, both $\mathrm{U(1)}$ and $\mathbb{Z}^T_2$ symmetry play nontrivial role.  It can be understood through Witten effect discussed in Refs.~\onlinecite{yegu2014,3d5}.   The TQFT description of this $\mathbb{Z}_2$ index is  given by a multi-component $b\wedge da $ in Sec. VI of Ref.~\onlinecite{yegu2014}.

\subsection{$\mathrm{U(1)}\times\mathbb{Z}_2^T$}
The complete classification of  bosonic SPT  phases with   $\mathrm{U(1)}\times \mathbb{Z}_2^T$ symmetry is given by $(\mathbb{Z}_2)^4$ \cite{Wencoho},  among which there are two $\mathbb{Z}_2$ indices are given by    bosonic SPT      phases in \ref{appendix_time_reversal}.   The remaining two $\mathbb{Z}_2$ indices  label irreducible  bosonic SPT  phases  with $\mathrm{U(1)}\times \mathbb{Z}_2^T$ symmetry.

 One $\mathbb{Z}_2$ index can be understood in the similar way to the Witten effect in bosonic topological insulators discussed in Refs.~\onlinecite{yegu2014,3d5}. The response theory is still given by $\Theta=2\pi$ $F\wedge F$ response action where the external gauge field $A$ here is ``spin gauge field'' that is pseudo-vector.  Under time-reversal transformation ``electric field'' $\mathbf{E}$ changes sign while ``magnetic field'' $\mathbf{B}$ not \cite{2d61}. The TQFT description of this $\mathbb{Z}_2$ index is   the same as the multi-component $b\wedge da $ in Ref.~\onlinecite{yegu2014} by just changing the time-reversal transformation of external $A$ from polar-like to pseudo-like transformations.

The other $\mathbb{Z}_2$ index is signalled by surface $\mathbb{Z}_2$ topological order where $e$ carries half charge (compared to the fundamental $\mathrm{U(1)}$ charge unit of  bulk bosons) and $m$ carries Kramers' doublet \cite{3d4}.  For the purpose of TQFT description,  we still start with the following two-component $b\wedge da$ theory with a coupling to external ``spin gauge field'' $A$:
\begin{align}
S=&\int i\frac{1}{2\pi} {b}^1\wedge \text{d} {a}^2 +i\int \frac{1}{2\pi} {b}^2\wedge \text{d}  {a}^1\nonumber\\
&+i\int \frac{2}{2\pi} {b}^2\wedge\text{d}{a}^2+S_{\rm coupling}\,.\label{newlagr}
\end{align}
It corresponds to $\frac{i}{2\pi}K^{IJ}\int b\wedge da$ with $K=\left(\begin{smallmatrix}
      0 & 1\\
        1 & 2
            \end{smallmatrix} \right)$.  $S_{\rm coupling}$ is given by: ($q_1,q_2\in\mathbb{Z}\,,F=dA$)
  \begin{align}
 S_{\rm coupling}=\int  \frac{iq_1}{2\pi}F \wedge d {a}^2 +\frac{iq_2}{2\pi}b^2\wedge dA \,.\label{eq:3dresponse}
\end{align}
Time-reversal transformation is defined as: ($I=1,2$; $i,j,\cdots=\hat{x},\hat{y},\hat{z}$)
\begin{align}
 \mathcal{T} a^I_0 \mathcal{T}^{-1}=a^I_0\,,& \mathcal{T} a^I_i \mathcal{T}^{-1}=-a^I_i\,,  \\
 \mathcal{T} b^I_{0i} \mathcal{T}^{-1}=-b^I_{0i}\,, &\mathcal{T} b^I_{ij} \mathcal{T}^{-1}=b^I_{ij}\,.
 \end{align}

  Integrating out $a^1$ leads to local flatness of $b^2$, which can be resolved by introducing a new 1-forma gauge field: $b^2=d\tilde{a}^2$.
Thus, the time-reversal transformation is given by:
 \begin{align}
 &\mathcal{T} \tilde{a}^2_0 \mathcal{T}^{-1}=-\tilde{a}^2_0\,, \mathcal{T} \tilde{a}^I_i \mathcal{T}^{-1}=\tilde{a}^2_i\,.\label{a_tilde}
 \end{align}

   The term $\sim  {b}^2\wedge \text{d}  {a}^2$ in Eq.~(\ref{newlagr}) provides  a surface Chern-Simons term:
\begin{align}
\mathcal{L}_{\partial}=i\frac{1}{\pi}\epsilon^{\mu\nu\lambda} \tilde{a}^2_\mu\partial_\nu  {a}^2_\lambda\label{eq:toriccode}
\end{align}
which can also be reformulated by introducing a matrix $K_{\partial}\xlongequal{\text{def.}}\left(\begin{smallmatrix}
      0 & 2\\
        2 & 0
            \end{smallmatrix} \right)$ in the standard convention of $K$-matrix Chern-Simons theory.\cite{Wenbook} $ {a}^2_\mu$ and $\tilde{a}^2_\mu$ form a 2-dimensional vector $( {a}^2_\mu, \tilde{a}^2_\mu)^T$. The ground state of Eq.~(\ref{eq:toriccode}) supports a $\mathbb{Z}_2$ topological order
 associated with four gapped quasiparticle excitations ($1,e,m, \varepsilon$).   By using $b^2=d\tilde{a}^2$, Eq.~(\ref{eq:3dresponse}) reduces to its surface counterpart:
$  \frac{q_1}{2\pi}\epsilon^{\mu\nu\lambda}A_{\mu}\partial_\nu  {a}^2_\lambda  +\frac{q_2}{2\pi}\epsilon^{\mu\nu\lambda}A_\mu \partial_\nu \tilde{a}^2_{\lambda}\,. $

Based on the Chern-Simons term in Eq.~(\ref{eq:toriccode}),  one may calculate the electric charge  carried by each quasiparticle: $Q_e=(q_1,q_2)(K_\partial)^{-1}(1,0)^T=\frac{q_2}{2}\,,\,Q_m=(q_1,q_2)(K_\partial)^{-1}(0,1)^T=\frac{q_1}{2}\,$.  Physically, both $e$ and $m$ quasiparticles can always attach trivial identity particles to change their charges by arbitrary integer so that $q_1$ and $q_2$ are integers mod $2$,  namely,
$
q_1\sim q_1+2\,,q_2\sim q_2+2\,.$
If  we consider $(q_1,q_2)=(2,1)$, then $Q_e=\frac{1}{2}\,,\,Q_m=1$ indicating that $e$ carries half-charge of $\mathrm{U(1)}$ symmetry. Since $\tilde{a}^2_\mu$ transforms as a pseudo-vector under time-reversal transformation defined in Eq.~(\ref{a_tilde}), its gauge charge, $m$ particle, is able to carry Kramers' doublet. This surface is what we need: $e$ carries half-charge while $m$ carries Kramers' doublet. We like to point out that the surface state cannot be realized on 2D plane alone unless time-reversal symmetry is broken. More concretely, let us investigate the coupling term ``$ \frac{q_2}{2\pi}\epsilon^{\mu\nu\lambda}A_{\mu}\partial_\nu  \tilde{a}^2_\lambda $ that changes sign under time-reversal transformation since $A$ is a pseudo-like vector:
$   \mathcal{T} A_0 \mathcal{T}^{-1}=- A_0\,, \mathcal{T} A_i \mathcal{T}^{-1}=A_i\,.
$  Therefore, this state necessarily break time-reversal on 2D plane. Since the sign change can be remedied by shifting $q_2$ through the identification ``$q_1\sim q_1+2\,,q_2\sim q_2+2\,$'',   the surface state is time-reversal invariant. Note that is the external gauge field is the usual electromagnetic field that is a polar-like vector, then, the symmetry group is $\mathrm{U(1)}\rtimes \mathbb{Z}^T_2$ and there are no obstruction for realization of the above surface state on 2D plane.

\end{document}